\documentclass[preprint2]{aastex}
\begin{document}

\title{\large{\rm{THE PULSATION MODE AND DISTANCE OF THE CEPHEID FF AQUILAE}}}

\author{\small D.~G. Turner$^1$, V.~V. Kovtyukh$^{2,3}$, R.~E. Luck$^4$, L.~N. Berdnikov$^{5,6}$}
\affil{$^1${\footnotesize Department of Astronomy and Physics, Saint Mary's University, Halifax, NS B3H 3C3, Canada.}}
\affil{$^2${\footnotesize Astronomical Observatory, Odessa National University, T. G. Shevchenko Park 65014 Odessa, Ukraine.}}
\affil{$^3${\footnotesize Isaac Newton Institute of Chile, Odessa Branch Ukraine.}}
\affil{$^4${\footnotesize Department of Astronomy, Case Western Reserve University, 10900 Euclid Avenue, Cleveland, OH 44106-7215, USA.}}
\affil{$^5${\footnotesize Moscow M.~V.~Lomonosov State University, Sternberg Astronomical Institute, Moscow 119992, Russia.}}
\affil{$^6${\footnotesize Isaac Newton Institute of Chile, Moscow Branch, Universitetskij Pr. 13, Moscow 119992, Russia.}}
\email{turner@ap.smu.ca}

\begin{abstract}
The determination of pulsation mode and distance for field Cepheids is a complicated problem best resolved by a luminosity estimate. For illustration a technique based on spectroscopic luminosity discrimination is applied to the $4^{\rm d}.47$ s-Cepheid FF Aql. Line ratios in high dispersion spectra of the variable yield values of $\langle M_V \rangle = -3.40 \pm 0.02$ s.e.~($\pm0.04$ s.d.), average effective temperature T$_{\rm eff} = 6195 \pm 24$ K, and intrinsic color $(\langle B \rangle - \langle V \rangle)_0 = +0.506 \pm 0.007$, corresponding to a reddening of $E_{B-V}=0.25\pm0.01$, or $E_{B-V}{\rm (B0)}=0.26\pm0.01$. The skewed light curve, intrinsic color, and luminosity of FF Aql are consistent with fundamental mode pulsation for a small amplitude classical Cepheid on the blue side of the instability strip, not a sinusoidal pulsator. A distance of $413\pm14$ pc is estimated from the Cepheid's angular diameter in conjunction with a mean radius of $\langle R \rangle=39.0\pm0.7\;R_{\odot}$ inferred from its luminosity and effective temperature. The dust extinction towards FF Aql is described by a ratio of total-to-selective extinction of $R_V=A_V/E(B-V)=3.16\pm0.34$ according to the star's apparent distance modulus.
\end{abstract}

\keywords{stars: variables: Cepheids---stars: fundamental parameters---stars: individual (FF Aql)---ISM: dust, extinction.}

\section{{\rm \footnotesize INTRODUCTION}}

A defining characteristic of most Cepheid variables is an asymmetric light curve with a rapid rise to maximum light followed by a slower decline to minimum 0.6--0.7 cycle later. For pulsation periods of $5^{\rm d}$ to $22^{\rm d}$ there is a superposed period-dependent secondary bump, the Hertzsprung progression \citep[see][]{tu12b}, that has enabled astronomers to establish diagnostics for the light curves of short-period Cepheids ($P < 10^{\rm d}$) from low-order Fourier series fits \citep{sl81}. Fundamental mode and overtone pulsators follow distinct trends in their Fourier parameters as a function of pulsation period that in principle allow identification of pulsation mode \citep*[e.g.,][]{ap86,an88,mp92,pp97}.

\begin{figure*}[!t]
\epsscale{0.85}
\plotone{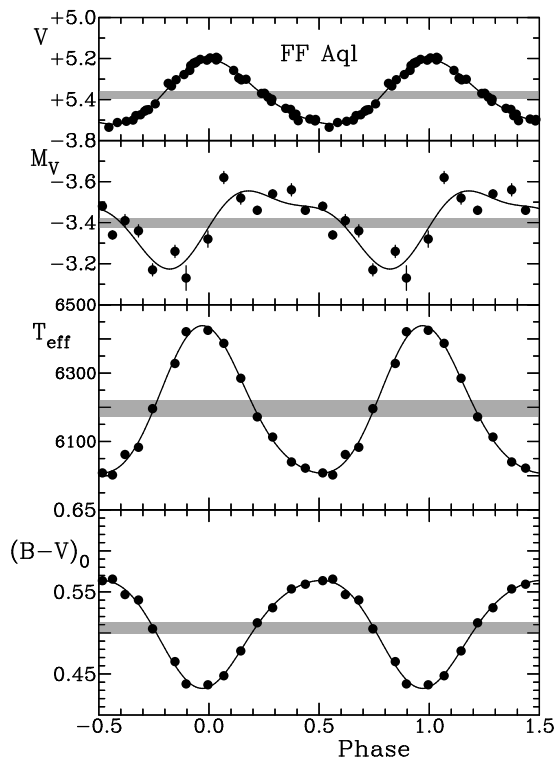}
\caption{\small{The phased variations in, from top to bottom, visual magnitude \citep{mb80}, absolute magnitude $M_V$, effective temperature T$_{\rm eff}$, and associated intrinsic color variations $(B-V)_0$ for FF Aql, the lower three from  the spectra analyzed here, with uncertainties in the data indicated. The gray relations represent the adopted mean values and the superposed curves are best-fitting Fourier series.}}
\label{fig1}
\end{figure*}

\begin{deluxetable}{@{\extracolsep{-2mm}}ccccccccccc}
\tabletypesize{\scriptsize}
\tablewidth{0pt}
\tablecaption{Spectroscopic Results for FF Aquilae \label{tab1}}
\tablehead{
\colhead{JD(obs)} &\colhead{Phase} &\colhead{$T_{\rm eff}$} &\colhead{$\pm$ s.d.} &\colhead{n}  &\colhead{$\pm$ s.e.} &\colhead{$M_V$} &\colhead{$\pm$ s.d.} &\colhead{n} &\colhead{$\pm$ s.e.} &\colhead{(B--V)$_0$} \\
& &\colhead{(K)} &\colhead{(K)} &  &\colhead{(K)} & & & & &} 
\startdata
2450672.747 &0.562 &6002 &81 &87 &9 &--3.34 &0.15 &67 &0.02 &0.566 \\
2450674.685 &0.995 &6425 &95 &72 &10 &--3.32 &0.30 &48 &0.04 &0.437 \\
2450675.687 &0.220 &6172 &77 &85 &8 &--3.46 &0.18 &70 &0.02 &0.512 \\
2450677.746 &0.680 &6083 &105 &68 &10 &--3.36 &0.18 &31 &0.03 &0.540 \\
2450678.712 &0.896 &6421 &66 &63 &10 &--3.13 &0.35 &39 &0.06 &0.438 \\
2450735.595 &0.619 &6062 &43 &90 &5 &--3.41 &0.23 &67 &0.03 &0.547 \\
2450736.608 &0.846 &6328 &61 &82 &7 &--3.26 &0.22 &67 &0.03 &0.465 \\
2450737.600 &0.067 &6387 &38 &61 &5 &--3.62 &0.27 &68 &0.03 &0.448 \\
2450738.592 &0.289 &6113 &49 &68 &6 &--3.54 &0.13 &67 &0.02 &0.531 \\
2450739.607 &0.516 &6008 &58 &70 &7 &--3.48 &0.17 &63 &0.02 &0.564 \\
2450740.625 &0.744 &6196 &78 &68 &9 &--3.17 &0.22 &70 &0.03 &0.505 \\
2451056.690 &0.437 &6022 &71 &58 &11 &--3.46 &0.16 &55 &0.02 &0.559 \\
2451095.619 &0.145 &6285 &53 &60 &7 &--3.52 &0.26 &65 &0.03 &0.478 \\
2451096.647 &0.374 &6040 &49 &68 &6 &--3.56 &0.20 &63 &0.03 &0.554 \\
\enddata
\end{deluxetable}

A complication arises from a small subset of Cepheids with periods $P < 7^{\rm d}$ that have small light amplitudes and light curves that are almost perfectly symmetric, making them very close matches to sine waves. Many such s-Cepheids appear to be overtone pulsators according to diagnostics that include first and second order Fourier terms \citep{mp92}, yet the meaning of such parameters is unclear when the light curve is a close match to a sine wave for which second order Fourier terms are undefined \citep[see][]{tu12b}. Is the Fourier fit providing a match to the Cepheid's true light variations or is it biased by random scatter in the observations? Such considerations are important given that s-Cepheids constitute a portion of the Cepheid demographic in all galaxies. Whether they are fundamental mode or overtone pulsators affects the luminosity expected for them from their periods of pulsation, and hence the inferred distance to a galaxy derived from the total Cepheid sample according to the Leavitt law.

A full characterization of individual Cepheids includes luminosity, which itself can separate fundamental mode pulsators from overtone Cepheids since the latter are typically $\sim0^{\rm m}.5$ more luminous than fundamental mode pulsators of identical pulsation period. Luminosity estimates require either an accurate knowledge of distance and extinction for individual Cepheids or a precise determination of their spectroscopic absolute magnitudes. Cluster Cepheids are useful for the former, provided their reddening is well established. It was noted by \citet{tm06}, for example, that the possible association of the s-Cepheids EU Tau, EV Sct, DX Gem, SZ Tau, BY Cas, V1334 Cyg, $\alpha$ UMi, BD Cas, QZ Nor, and V1726 Cyg with the clusters Alessi 90, NGC 6664, Alessi J0652.6+1439, NGC 1647, NGC 663, Dolidze 45, Harrington 1, King 13, NGC 6067, and Platais 1, respectively, imply a mix of pulsation modes for the pulsators: fundamental mode pulsation for EU Tau, DX Gem, BY Cas, $\alpha$ UMi, and V1726 Cyg, and overtone pulsation for EV Sct, SZ Tau, V1334 Cyg, QZ Nor, and BD Cas.

The technique of determining Cepheid absolute magnitudes spectroscopically by averaging numerous line ratios relating neutral to ionized species of iron group elements in their spectra has been developed by Kovtyukh and collaborators \citep*{ko08,ko10,ko12a}. In practice a single line ratio can specify absolute magnitude $M_V$ to $\pm0^{\rm m}.26$ for FG supergiants. Application to all potential luminosity-sensitive line ratios in an observed spectrum can reduce the overall uncertainty in absolute magnitude by $1/n^{\frac{1}{2}}$ according to the number ({\it n}) used, provided the uncertainties in each estimate are governed by random errors of measurement. There is no evidence otherwise. The technique yields Cepheid luminosities of greater precision ($\sim1\%$) than other methods, and is fairly robust. An application to the s-Cepheids V1334 Cyg, V440 Per, and V636 Cas by \citet{ko12b} confirmed the overtone nature of V1334 Cyg inferred from cluster membership \citep{tm06} and implied fundamental mode pulsation for V440 Per and V636 Cas. A subsequent application to $\alpha$ UMi \citep{te13} confirmed the case for fundamental mode pulsation in the Cepheid, despite a conflict with results implied by the star's Hipparcos parallax \citep{vl13}. Of the dozen s-Cepheids studied to date, 7 (58\%) appear likely to be fundamental mode pulsators while only 5 (42\%) are overtone pulsators. Apparently a sinusoidal light curve is not a characteristic of a specific pulsation mode.

This paper examines the case of the $4^{\rm d}.47$ s-Cepheid FF Aql, which presents its own challenges. It is not associated with an open cluster, according to previous surveys, and conclusions about its pulsation mode differ  according to which measured parallax for the star is adopted. The HST parallax for FF Aql \citep{bn07} implies fundamental mode pulsation, while its Hipparcos parallax \citep{vl07} matches expectations for overtone pulsation. Alternatively, its rate of period change is consistent with fundamental mode pulsation \citep{tu10}, although the match is not ideal. Possible contamination of the star's photometry by nearby and unresolved companions \citep{ev90,ue93} is a problem, but is of only minor concern for spectroscopic observations. As demonstrated here, FF Aql is one case where the spectroscopic method of deducing luminosity may be ideally suited to resolving uncertainties about a Cepheid's pulsation mode and evolutionary status.

\section{{\rm \footnotesize OBSERVATIONS AND DATA REDUCTIONS}}
The observations used for the present study consist of the McDonald Observatory Sandford echelle spectrograph observations for FF Aql included in the earlier study of s-Cepheids by \citet{lu08}, further analyzed to establish absolute magnitude estimates $M_V$ for the Cepheid as a function of pulsation phase. Correct phasing of the observations was made using the ephemeris of \citet{bp94}, as done by \citet{lu08}. The Cepheid exhibits a slow evolutionary period increase superposed upon its orbital O--C changes \citep{bp94,tu10}, although a study to separate such effects by \citet{be97} left the matter unresolved. FF Aql is a spectroscopic binary with an orbital period of about 3.92 years \citep{ab59,ev90,go95}, making it difficult to separate O--C changes arising from evolution from scatter introduced by the Cepheid's orbital motion \citep{bp94}.

The results of the luminosity calculations for FF Aql derived from the observed spectra are summarized in Table~\ref{tab1} along with the {\it T}$_{\rm eff}$ changes found earlier by \citet{lu08}. The latter have been converted into equivalent variations in intrinsic $(B-V)_0$ color using the relationship between the two derived by \citet{gr92}. Uncertainties in the inferred values of {\it T}$_{\rm eff}$ and $M_V$ are included in the table, and the results are presented graphically as a function of pulsation phase in Fig.~\ref{fig1} along with the Cepheid and its unresolved companion's visual brightness variations from \citet{mb80}.

\section{{\rm \footnotesize RESULTS}}

Initially the observed variations in visual magnitude, absolute magnitude, effective temperature, and intrinsic color were matched to best fitting sine waves, but that yielded sizable systematic residuals in most cases. Low order Fourier series of fairly similar morphology for each parameter produced much smaller residuals, and are the fitted relations shown in Fig.~\ref{fig1}. The inferred mean values are: $\langle V \rangle=5.38\pm0.02$, $\langle M_V \rangle=-3.40\pm0.02$ s.e.~($\pm0.04$ s.d.), $\langle T_{\rm eff} \rangle = 6195 \pm24$ K, and mean intrinsic color $\langle (B-V)_0 \rangle=0.506\pm0.007$. The observed mean color of FF Aql is $\langle B-V \rangle=0.755$ \citep{be07}, implying a reddening of $E_{B-V}=0.25\pm0.01$. The corresponding color excess for a B0 star observed through the same amount of extinction is $E_{B-V}{\rm (B0)}=0.26\pm0.01$ according to the relationship of \citet{fe63}.

The best fitting curve for absolute magnitude closely resembles the skewed light curves of standard Cepheids, with a rapid rise to maximum followed by a slow decline to minimum. It is less noticeable in photometric observations for the Cepheid \citep{mb80}, presumably because of contamination by the star's companions. FF Aql is recognized to have a nearby optical companion $5^{\rm m}.8$ fainter \citep{ue93}, a close speckle companion perhaps $2^{\rm m}$ to $4^{\rm m}$ fainter \citep{mc87,mc89,ev90}, and an unresolved spectroscopic companion estimated to be $\sim 6^{\rm m}$ fainter \citep{ev90}. Although the net effect on the Cepheid's observed brightness variations from aperture photometry is small, the constant presence of light contamination in the observations and potential color effects arising from the different temperatures of the companions likely smooth out the skewness in the Cepheid's visual light variations (Fig.~\ref{fig1}, top).

The mean absolute magnitude inferred for FF Aql of $\langle M_V \rangle = -3.40 \pm0.02$ is consistent with fundamental mode pulsation for a $4^{\rm d}.47$ Cepheid. The corresponding Fourier parameters of $\phi_{21}=5.69\pm0.66$, $\phi_{31}=4.46\pm0.94$, R$_{21}=0.37\pm0.20$, and R$_{31}=0.27\pm0.20$ are also generally consistent with fundamental mode pulsation, as noted for LMC and M31 Cepheids \citep{be95,vi07} as well as Galactic Cepheids \citep{an88,za00,st11}, but the match is not ideal in $\phi_{21}$ and $\phi_{31}$, presumably because of scatter in the data. In any case, the skewed nature of the absolute magnitude curve for FF Aql in conjunction with its skewed radial velocity variations \citep{ev90} and the similarly skewed variations in $T_{\rm eff}$ and $(B-V)_0$ indicate that it is not a true s-Cepheid, so does not belong in that category.

\begin{figure}[!t]
\epsscale{0.85}
\plotone{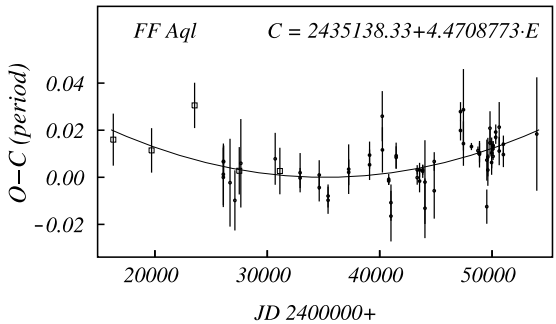}
\caption{\small{Observed minus predicted times of light maximum (O--C) for FF Aql, in units of fractions of its pulsation period, with associated uncertainties. The long-term trend indicated corresponds to $\log {\dot{P}}=-1.153$, in units of s yr$^{-1}$.}}
\label{fig2}
\end{figure}

The O--C variations for FF Aql tabulated by \citet{bp94} and \citet{be97}, and updated by additional times of light maximum derived from ASAS observations and Harvard photographic plates \citep{bt13}, is shown in Fig.~\ref{fig2}. The slow period increase of $+0.0703 \pm 0.0160$ s yr$^{-1}$ \citep{bp94,bt13} for FF Aql was shown previously to be consistent with the rate expected from stellar evolutionary models for a Cepheid pulsating in the fundamental mode in the third crossing of the instability strip \citep{tu10}. The larger reddening found here for the Cepheid confirms that conclusion, since the blue intrinsic color for FF Aql relative to other $4^{\rm d}.47$ Cepheids places it solidly on the blue edge of the instability strip \citep*[see][]{te06,tu10}, a region dominated by small amplitude variables \citep[see papers cited by][]{te06}. The observed rate of period increase ($\log {\dot{P}}=-1.153$) is also consistent with expectations for a Cepheid on the hot edge of the instability strip, since it is slightly larger than values derived for other third crossing Cepheids (i.e., slow period increases). Cepheids of a given period on the hot edge of the instability strip are more massive and are evolving faster than those near strip center \citep[see discussion by][]{te06}, so their rates of period change are systematically higher. Fundamental mode pulsation for FF Aql is therefore consistent with its derived parameters.

\section{{\rm \footnotesize THE DISTANCE TO FF AQL}}
A variety of methods enable one to establish the distance to FF Aql. As noted in \S1, there is a discrepancy between the results obtained from trigonometric parallaxes. The distance to FF Aql implied by the star's HST parallax \citep{bn07} is $356\pm23$ pc, whereas its Hipparcos parallax \citep{vl07} yields a distance of $474\pm74$ pc, results that are discordant within the cited uncertainties. The difference accounts for previous questions about the pulsation mode of FF Aql \citep[see][]{tu10}. The infrared surface brightness (IRSB) technique yields a distance of $370\pm10$ pc \citep{st11}, a higher precision result consistent with the HST parallax.

The mean luminosity for FF Aql derived here can be used to establish the distance to the Cepheid independently. A first step is to derive the mean radius of FF Aql from its luminosity and effective temperature relative to solar parameters \citep[see][]{tb02}. That yields a mean radius of $\langle R \rangle=39.0\pm0.7\;R_{\odot}$, a value larger than that obtained by \citet{ga12} using the interferometric Baade-Wesselink method, yet in better agreement with the period-radius relations they used for comparison as well as with the period-radius relations of \citet{tb02}, \citet{te10}, and \citet{mo11}. The limb-darkened angular diameter of $\theta_{\rm LD}=0.878\pm0.013$ mas found for FF Aql by \citet{ga12} from near-infrared interferometry then combines with the star's estimated mean radius to yield a distance of $413\pm14$ pc to FF Aql, a value lying roughly midway between the two estimates from the star's trigonometric parallax and larger than the value obtained by the IRSB method.

A weighted mean of the four semi-independent estimates yields a distance of $388\pm8$ pc, while an unweighted average is $398\pm50$ (s.d.) pc, but that assumes no sources of systematic error in each, which may not be valid. The origin of the discrepancy in the spectroscopic distance relative to the values associated with the HST and Hipparcos parallaxes is uncertain, although the presence along the line of sight to FF Aql of a moderately bright companion with a changing optical separation relative to the Cepheid \citep{mc87,mc89} may introduce bias into parallax estimates tied to measurements of the star's photocenter. The photometric contamination of the companions to FF Aql on its brightness and colors may also account for the discrepancy relative to the IRSB technique.

\begin{deluxetable}{@{\extracolsep{+2mm}}ccc}
\tabletypesize{\small}
\tablewidth{0pt}
\tablecaption{Photometric Properties of FF Aquilae \label{tab2}}
\tablehead{
\colhead{Parameter} &\colhead{FF Aql} &\colhead{Uncertainty} }
\startdata
$\langle M_V \rangle$ &--3.40 &$\pm0.02$ \\
$(B-V)_0$ &+0.506 &$\pm0.007$ \\
T$_{\rm eff}$ (K) &6195 &$\pm24$ \\
$E_{B-V}$ &0.25 &$\pm0.01$ \\
R$_{21}$ &0.37 &$\pm0.20$ \\
R$_{31}$ &0.27 &$\pm0.20$ \\
$\phi_{21}$ &5.69 &$\pm0.66$ \\
$\phi_{31}$ &4.46 &$\pm0.94$ \\
$\dot{P}$ (s yr$^{-1}$) &+0.0703 &$\pm0.0160$ \\
$\langle R \rangle /R_{\odot}$ &39.0 &$\pm0.7$ \\
d (pc) &413 &$\pm14$ \\
$R_V$ &3.16 &$\pm0.34$ \\
\enddata
\end{deluxetable}

It is possible to go further to determine a value for the ratio of total-to-selective absorption $R_V=A_V/E(B-V)$ that applies to dust along the line of sight to FF Aql. That requires an estimate of the apparent distance modulus for the Cepheid, which is somewhat difficult given uncertainties about the amount of light contamination inherent to the optical photometry. The optical and spectroscopic companions are faint enough \citep{ev90} to affect the total light of the system by only $0^{\rm m}.01$, but contamination by the interferometric companion may be responsible for $0^{\rm m}.04$ to $0^{\rm m}.22$ of the total visual light from FF Aql \citep[see][]{ev90}. Adoption of a simple mean of the latter values leads to an apparent distance modulus of $V-M_V=8.91\pm0.07$, which yields an estimated value of $R_V=3.16\pm0.34$ for the extinction properties of the dust towards the Cepheid.

The region of Aquila is not well studied in terms of extinction along the line of sight. Open clusters lying within $\sim10^{\circ}$ of FF Aql yield a mean value of $R_V=3.18\pm0.18$ from variable-extinction studies \citep{tu76}, while a value of $R_V=3.0$ was adopted by \citet{fb84} in his study of the distribution of OB stars in this direction. The value of $R_V$ obtained here from the spectroscopic luminosity and effective temperature of FF Aql is therefore consistent with previous studies of the dust extinction in Aquila. Note that adoption of the HST parallax for FF Aql would yield a value of $R_V\simeq4$ for the extinction, while adoption of the Hipparcos parallax would yield a value of $R_V\simeq2$, albeit with an uncertainty that permits agreement with the present result. The more typical value ($R_V\simeq3$) following from the present results appears to provide further substantiation for the robust nature of the spectroscopic luminosity technique.

\begin{figure}[!t]
\epsscale{0.95}
\plotone{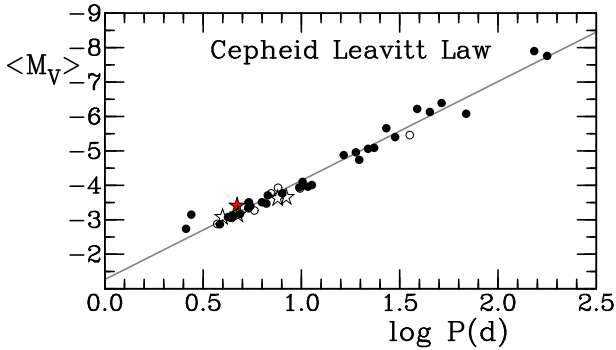}
\caption{\small{The dependence of absolute visual magnitude on pulsation period for cluster Cepheids and Cepheid-like objects (filled circles), HST Cepheids (open circles), and spectroscopic parallax Cepheids (stars). FF Aql is denoted by the red-filled star, and the gray relation is a least squares fit for cluster Cepheids.}}
\label{fig3}
\end{figure}

\section{{\rm \footnotesize SUMMARY}}

A study of phase-dependent spectroscopic variations in absolute visual magnitude $M_V$ and effective temperature T$_{\rm eff}$ for the $4^{\rm d}.47$ Cepheid FF Aql reveals a skewed nature for its uncontaminated light curve (Fig.~\ref{fig1}) that implies it is not a sinusoidal s-Cepheid. That is supported by Fourier parameters inferred from the Cepheid's absolute magnitude variations (Table~\ref{tab2}), which differ from similar values derived from optical photometry \citep{st11}, presumably because of contamination by the Cepheid's three close companions. Although \citet{st11} assumed that FF Aql is a s-Cepheid, the present results contradict such a conclusion.

Observed variations in effective temperature have been linked to phased variations in intrinsic color $(B-V)_0$ to derive a reddening of $E_{B-V}=0.25\pm0.01$ ($E_{B-V}{\rm (B0)}=0.26\pm0.01$) for FF Aql (Table~\ref{tab2}). An identical reddening was found by \citet{tu87} from {\it JHK} photometry, but lies on the high end of values (0.191 to 0.250) cited by \citet{tu10}, which include an estimate of 0.224 by \citet{ko08} from the same spectra. It is larger than the reddening of 0.196 cited by \citet{st11}. Contamination of optical photometry by the Cepheid's companions may be responsible for the differences.

This study yields a mean radius and effective temperature for FF Aql (Table~\ref{tab2}) that have been combined with the star's mean angular diameter \citep{ga12} to derive a distance of $413\pm14$ pc, a value lying within the range of estimates from the star's HST \citep{bn07} and Hipparcos \citep{vl07} parallaxes, as well as its IRSB method distance \citep{st11}, but falling outside the uncertainty estimates for all but the Hipparcos result. The resulting extinction along the line of sight to the Cepheid derived by spectrophotometric means appears to be relatively normal ($R_V=3.16\pm0.34$). The dust towards $l=49^{\circ}.21$ and $b=+6^{\circ}.36$ therefore appears to share similar properties to most other sight lines in the Galaxy.

The spectroscopic luminosity for FF~Aql derived here is compared in Fig.~\ref{fig3} with other best estimates of luminosity for Galactic Cepheids and Cepheid-like objects established by spectroscopic means \citep{ko12b,te13}, open cluster membership \citep{te09,tu10,te10,me11,me12a,me12b,te12a,te12b}, including new unpublished results for for S~Vul, and HST parallaxes \citep{bn07}. The result for FF~Aql in the last source was excluded in favor of the present spectroscopic estimate, and the HST luminosity for $\ell$~Car cited by \citet{tu10} was adjusted to account for an anomalous extinction law in Carina described by $R_V=A_V/E(B-V)\simeq4$ rather than $R_V\simeq3$ \citep{tu12a,ca13}. The periods of recognized overtone pulsators were also adjusted to their equivalent fundamental mode values. The location of FF~Aql relative to fundamental mode pulsation for other calibrating Cepheids in the $PL$ plane of Fig.~\ref{fig3} coincides closely with expectations and confirms the previous mode identification. FF~Aql is slightly overluminous (smaller $M_V$) relative to other calibrating Cepheids of comparable period ($4^{\rm d}.4$) because it lies towards the hot edge of the instability strip.
\scriptsize{}

\end{document}